\documentclass[doublecol]{epl2}
\usepackage{graphicx}

\title{Detection of fractional solitons in quantum spin Hall systems}
\shorttitle{High temperature STM detection of Wigner molecules in carbon nanotubes} 

\author{C. Fleckenstein \and N. Traverso Ziani \and B. Trauzettel}
\shortauthor{C. Fleckenstein \etal}

\institute{
  
  Institute for Theoretical Physics and Astrophysics, University of W{\"u}rzburg, D-97074 W{\"u}rzburg, Germany
}

\abstract{We propose two experimental setups that allow for the implementation and the detection of fractional solitons of the Goldstone-Wilczek type. The first setup is based on two magnetic barriers at the edge of a quantum spin Hall system for generating the fractional soliton. If then a quantum point contact is created with the other edge, the linear conductance shows evidence of the fractional soliton. The second setup consists of a single magnetic barrier covering both edges and implementing a long quantum point contact. In this case, the fractional soliton can unambiguously be detected as a dip in the conductance without the need to control the magnetization of the barrier.}

\begin{document}

\maketitle \section{Introduction}
Solitons with fractional charge represent a long standing bridge between particle physics and condensed matter physics\cite{revwilc}. They usually emerge in noninteracting systems that can be described by a Dirac equation with a non-uniform mass. The first and most prominent examples are the Jackiw-Rebbi model\cite{jr} and the Su Schrieffer Heeger (SSH)\cite{ssh} model for polyacetylene. In those two models, which are at low energy equivalent, domain walls/mass kinks carry a charge which is half of the electronic charge. The appearance of the half charge can be understood by symmetry reasons: any regularization scheme that allows us to calculate the number of fermions in Jackiw-Rebbi like systems predicts that, in the presence of chiral symmetry, non-degenerate zero energy states necessarily carry half-integer charge\cite{bell,rajaraman,overlapping}. Moreover, this fractional charge is a 'stable quantum number'\cite{overlapping,chiralanomaly,stable}. Shortly after the prediction of the half charge characterizing the Jackiw-Rebbi model, Goldstone and Wilczek\cite{fractional} proposed a more general model supporting solitons which can carry any fractional charge. Surprisingly, even if chiral symmetry is broken, the fractional charge characterizing the Goldstone-Wilczek model is a stable quantum number\cite{stable,chiralanomaly}. Experimental proposals for the realization and the detection of fractional solitons have been lacking until very recently. In fact, already the simplest case, the SSH model, poses limitations that are difficult to overcome. While it is possible to measure the bare existence of the solitons\cite{sshcold}, the spin degeneracy makes the detection of their half charge impossible.\\
The scenario has drastically changed with the advent of two-dimensional topological insulators\cite{ti1,ti2,ti3,ti4,ti5,ti6,ti7,ti8}. These materials are characterized by the presence of metallic edge states circulating around an insulating bulk. The metallic edge states exhibit perfect spin-momentum locking. This means that electrons moving in opposite directions have opposite spin projection. The possibilities opened by this property are countless. The protection against backscattering makes topological insulators potential candidates in electronic devices and their peculiar spin structure let us envision novel applications in spintronics\cite{spin1,spin2,spin3}. Moreover, the recently achieved possibility to induce superconductivity in topological insulators\cite{sup1,sup2,sup3} by the proximity effect makes them perfect candidates for superconducting spintronics\cite{sspin1,sspin2,sspin3,sspin4} and a valid alternative to spin-orbit coupled quantum wires\cite{qw1,qw2,qw3,qw4,qw5,qw6,qw7,qw8,qw9,qw10,qw11} in topological quantum computation, through the realization of Majorana fermions\cite{m1,m2,m3,m4,m5,m6,felix1} and parafermions\cite{pf1,pf2,pf3,pf4,pf5}.\\
As far as the appearance of fractional charges is concerned, it was shown\cite{fracnat} that two magnetic barriers with in-plane magnetization with an angular difference $\theta$ are able to trap a fractional charge $Q_F=\theta/2\pi$ (in units of the electron charge), see Fig.1$(a)$ for a schematic. This behaviour directly related to the Goldstone Wilczek solitons. These fractional excitations have been shown to be connected to a real space analogue of the chiral anomaly\cite{chiralanomaly,prb17}, and they are robust with respect to interactions\cite{giacomo1,giacomo2}. Strikingly, the case $\theta=1/2$ can also be realized by interaction induced spontaneous symmetry breaking, corresponding to fractional Wigner crystallization\cite{fracnat,mac,prl}.\\
As far as the experimental detection of the fractional charge is concerned, up to now the methods proposed have been proven to be difficult to implement. The original proposal has been based on measuring the shift of the conductance peaks of the linear conductance, characterizing a single edge with two magnetic barriers, as the angle of one magnetic barrier is varied\cite{fracnat}. There are both technical and conceptual problems related to this proposal. The technical problem is that it is difficult to implement and control the magnetic barriers on the edge of the quantum spin Hall system. The conceptual problem is that the fractional charge is a well defined quantity only in the limit of infinite magnetic barrier strength.\\
In this Letter, we propose a new scheme (see Fig.1$(b)$). The magnetic barriers are implemented in one of the two edges and, with respect to the quantum dot they define, a point contact with the second edge is formed. The linear conductance measured in the lower edge is then sensitive to the presence of the fractional soliton in the dot on the upper edge. In order to overcome the technical difficulty of implanting and rotating two magnetic barriers, we also propose another setup supporting fractional charge $1/2$. It consists of a single magnetic barrier covering both edges, and a quantum point contact (see Fig.1$(c)$). The fractional charge is then again measured as a feature in the linear conductance, as we describe in detail below.\\
The outline of the Letter is the following. We start by describing the system and the more traditional scheme for the detection of the fractional solitons. We then present our results on  the conductance in the system consisting of two magnetic barriers and a quantum point contact. Finally, we discuss the structure that only needs one barrier for trapping the half charge and show how the fractional soliton can be identified.
\begin{figure}
	\begin{center}
		\includegraphics[width=1\linewidth]{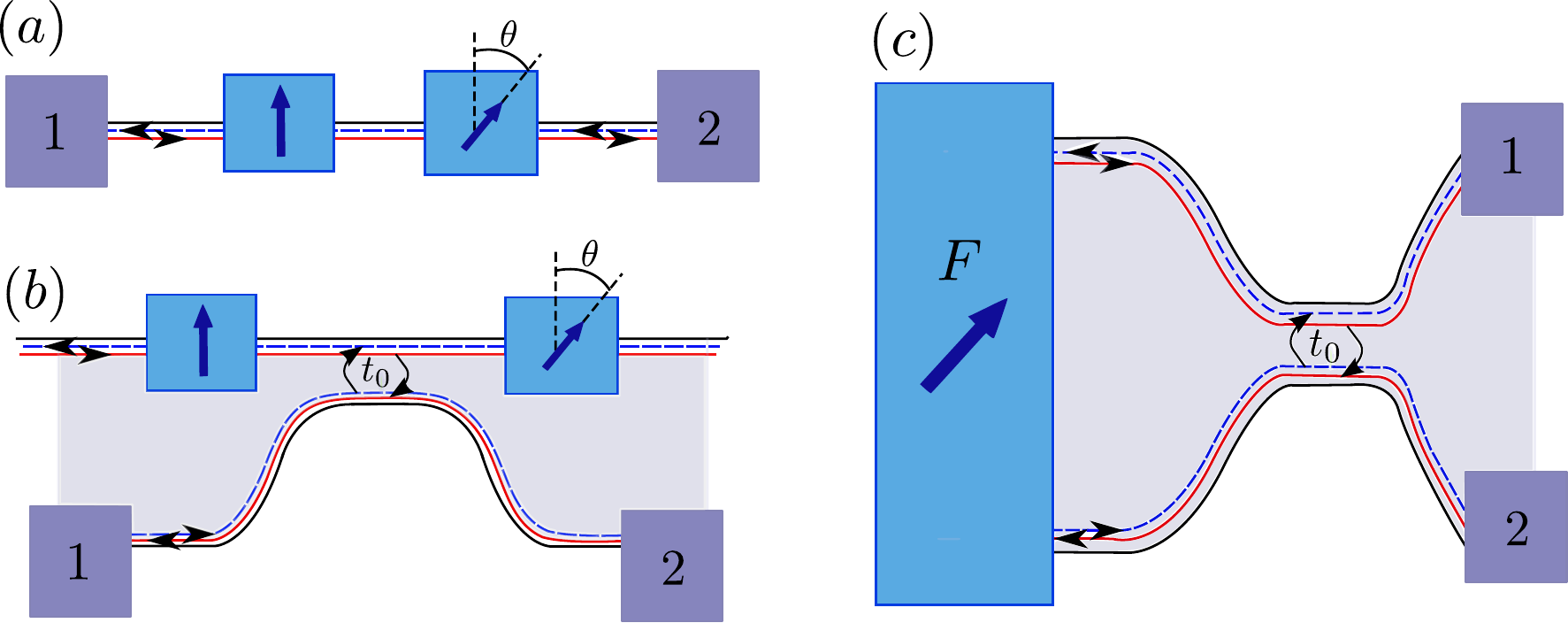}
		\caption{Schematic of the three setups. Red/blue lines correspond to spin up/down electrons. The light blue rectangular boxes represent the magnetic barriers and the purple ones the leads employed for transport.}
		\label{fig:e_B_mu3}
	\end{center}
\end{figure}
\section{Effective model}
\label{sec:isolated}
The setup presented in this section (Fig.1$(a)$) is not new\cite{fracnat}, but it represents the fundamental building block of the following new proposals so it is worth to recap it. We consider an infinite edge of a two-dimensional topological insulator with two magnetic barriers with in-plane magnetization. The corresponding Hamiltonian $H_1$ reads ($\hbar=1$)
\begin{equation}
H_1=\int_{-\infty}^\infty dx\left[ \Psi^\dagger (x) (-iv_F\sigma_z\partial_x) \Psi(x)+ \mathcal{H}_{M,1}(x)\right],
\end{equation}
where $\Psi(x)$ is a two component fermionic spinor, $v_F$ the Fermi velocity, $\sigma_z$ the third Pauli matrix (in the usual representation) and
\begin{eqnarray}
\mathcal{H}_{M,1}(x)=\Psi^\dag(x)M[\sigma_x Y(x+L)Y(-x)+\nonumber\\
(\cos \theta\, \sigma_x+\sin\theta \,\sigma_y )Y(x-d)Y(L+d-x) ]\Psi(x).
\end{eqnarray}
Here, $M$ parametrizes the strength of the barriers, $L$ and $d$ are the length of the barriers and the length of the dot, respectively, $\sigma_x$ and $\sigma_y$ are the $x$ and $y$ Pauli matrices and $Y(x)$ is the Heaviside function. In the limit $ML/v_F\rightarrow\infty$, a fractional charge $Q_F=\theta/2\pi$ is trapped between the two barriers, which hence define a particular quantum dot. The presence of the fractional charge has a direct consequence on the energy

 levels $\epsilon^{(1)}_n(\theta)$ in the quantum dot, which are given by
\begin{equation}
\epsilon^{(1)}_n(\theta)=\frac{v_F\pi}{d}\left(n-\frac{1}{2}+\frac{\theta}{2\pi}\right)
\end{equation}
with $n$ integer.
\begin{figure}
	\begin{center}
		\includegraphics[width=1\linewidth]{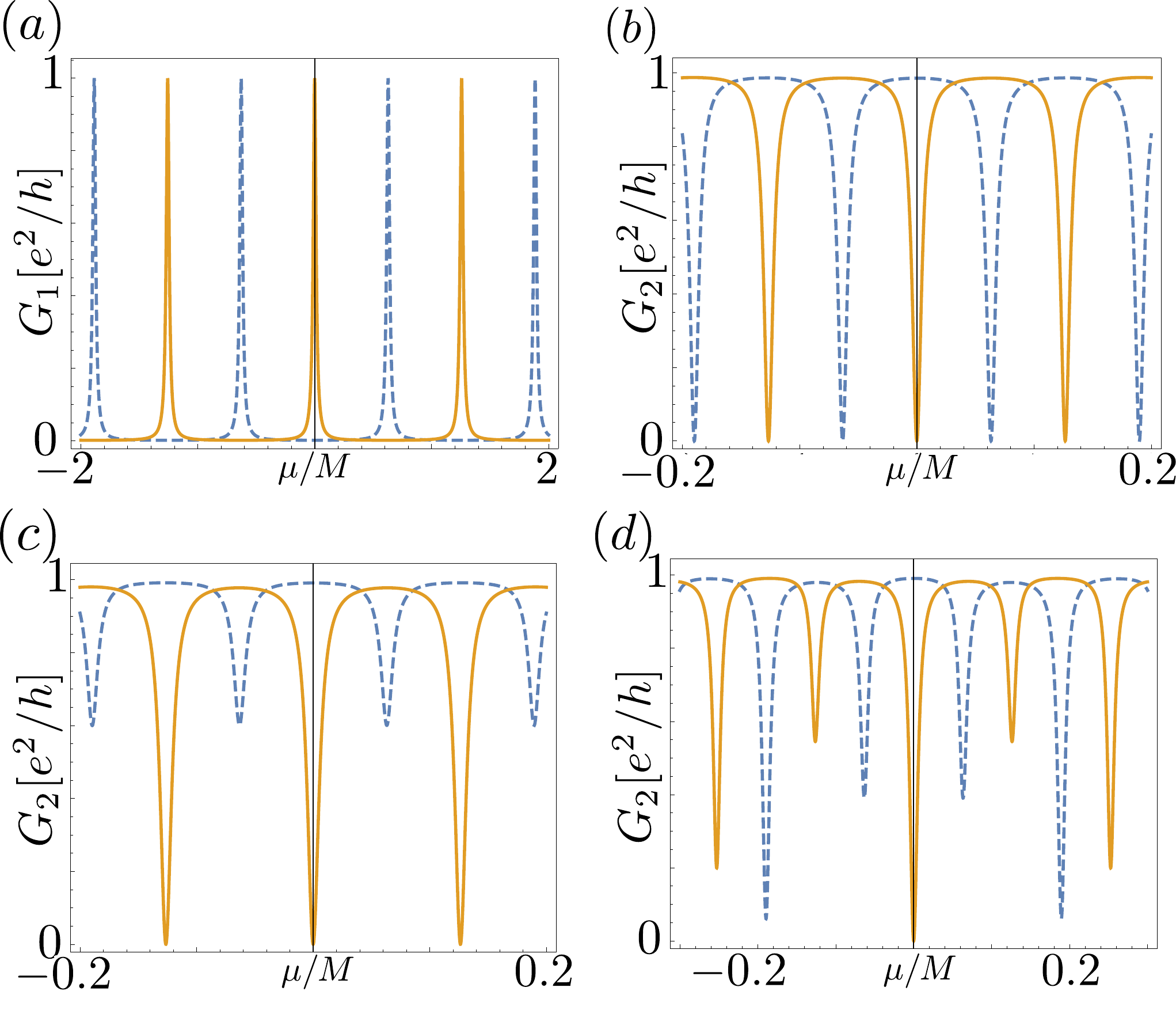}
		\caption{$(a)$ $G_1$ as a function of the gate voltage for $M=2v_F/L$, $d=1.25L$ and $\theta=\pi$ (orange solid line) and $\theta=0$ (blue dashed line)  $(b)$ $G_2$ for as a function of the gate voltage for $M=20v_F/L$, $d=1.25L$, $L_1=0.375L$, $L_2=0.625L$, $t_0=2v_F/L$, $t_s=0$ and $\theta=\pi$ (orange solid line) and $\theta=0$ (blue dashed line). $(c)$ same as in $(b)$ but for $t_s=v_F/L$. $(d)$ Same as in $c$ but for $L_1=0.275L$, $L_2=0.525L$.}
		\label{fig:e_B_mu}
	\end{center}
\end{figure}
The first proposed scheme for the detection of the fractional soliton exploits this dependence of the energy on the fractional excitation through $\theta$\cite{fracnat}. It consists of considering the linear conductance $G_1$ of the system when a small bias is applied between the parts of the helical edge left and right of the dot. Such quantity can be easily obtained by calculating the scattering matrix of the system\cite{sm1,sm2,sm3}. As shown in Fig.2$(a)$, as the magnetization direction of the impurities is flipped from parallel to antiparallel, the linear conductance peaks as a function of the gate voltage applied to the quantum dot (and hence of the chemical potential $\mu$ in the dot) are shifted by half of their period. This behaviour is a consequence of the presence of the fractional charge 1/2. There are two main difficulties of this proposal. In order to have peaks in the conductance with a significant amplitude, the magnetic barriers should allow for some hybridization between the states in the quantum dot and the states outside the quantum dot, meaning that the fractional soliton inside the dot hybridizes with the fractional, non-localized antisoliton outside the quantum dot (the total charge is then integer), which makes it difficult to argue for the existence of fractional charge in this setup\cite{chiralanomaly}. Moreover, the detection scheme relies on the ability to not only implant two magnetic barriers on the helical edge, but to be also able to flip the magnetization of one of the two as well. In the next section, we will overcome the first challenge.
\section{Detection by a point contact}
Now, we address the setup shown in Fig.1$(b)$. The total Hamiltonian $H_2$ is given by
\begin{equation}
H_2=H_1+\int_{-\infty}^{\infty}dx \chi^\dagger (x) (iv_F\sigma_z\partial_x) \chi (x)+H_{T,1},
\end{equation}
where, $\chi(x)$ is the Fermi spinor on the edge opposite to the one where the fractional charge is located. We will refer to the edge related to the field $\Psi(x)/\chi(x)$ as the upper/lower edge, respectively. Note that the helicity is opposite in the two edges. The Hamiltonian $H_{T,1}$ describes the point contact between the edges. Explicitly,
\begin{equation}
H_{T,1}=t_0\int_{L_1}^{L_2} \Psi^\dagger(x) I_{2\mathrm{x}2}\chi(x)+\mathrm{H.c.},
\end{equation}
where, $t_0$ parametrizes the strength of the tunneling amplitude and $L_1$ and $L_2$ indicate the location and the length of the point contact. The constraints $0<L_1<L_2<d$ are assumed to hold. The scheme for the detection of the energy levels of the quantum dot, and hence of the fractional soliton, is the following one: The linear conductance in the lower edge is measured as a function of the gate voltage applied to the quantum dot. When the gate voltage in the dot is chosen such that an energy level in the dot is on resonance for transport, there is a peak in the backscattered current at the lower edge. Equivalent information is conveyed by the nonlinear conductance at fixed gate potential in the quantum dot. Similar to the previous section, the key issue for the detection of the fractional soliton is that the peaks get shifted when the magnetization direction of one of the two barriers is rotated. The advantage of this setup, as compared to Fig.1$(a)$, is that the magnetic barriers do not need to have any transparency in order to show a significant conductance modulation atthe lower edge. Once the measurement is performed by means of the quantum point contact, the parameter $t_0$ can be then reduced for example by gating, leaving a stable fractional charge in the quantum dot behind. The results, obtained by means of a simple scattering matrix approach, are shown in Fig.2$(b)$. Here, the conductance $G_2$ of the lower edge is shown as a function of the chemical potential in the dot for parallel and antiparallel magnetization of the impurities. In the conductance calculation, we have assumed that the gate voltage also couples to the lower edge at the location of the quantum point contact. An important issue is that the mechanism we propose is crucially based on the breaking of time reversal symmetry in the edge containing the quantum dot. Furthermore, it is robust under the experimentally relevant case of broken axial spin symmetry\cite{helical1,generic1,generic2,generic3,prl}. The main implication of such a breaking, in the absence of electron-electron interactions, is to add a spin-flipping forward scattering term $H_{s}$ to the Hamiltonian $H_2$. Explicitly
\begin{equation}
H_s=t_s\int_{L_1}^{L_2}\Psi^\dag(x)\sigma_x\chi(x)+\mathrm{H.c.}.
\end{equation}
The result for $G_2$ in the presence of $H_s$ is shown in Fig.2$(c)$. Although quantitative differences are present, the qualitative behaviour due to the presence of the fractionally charged solitons is clearly visible. Importantly, the results do not qualitatively depend on the position of the quantum point contact (see Fig.2$(d)$).\\
In the next section, we present a setup where no rotation of the magnetic barriers is needed any more, hence, overcoming the main experimental difficulty affecting the schemes proposed so far.
\section{Fractional charge with a single barrier}
We now come to the most relevant part of the Letter. The system analysed in this section, see Fig.1$(c)$ for a schematic, consists of a magnetic barrier covering both edges of a two-dimensional topological insulator and an extended point contact between them. A similar setup, with a superconductor instead of a ferromagnetic barrier, has recently been proposed for the detection of Majorana fermions\cite{bernevig}. Explicitly, the Hamiltonian $H_3$ reads
\begin{equation}
H_{3}=H_{K,3}+H_{M_3}+H_{T,1}.
\end{equation}
Here, the constraints in $H_{T,1}$ are $0<L_1<L_2$ and
\begin{eqnarray}
H_{K,3}=v_F\int_{-\infty}^\infty dx\left[ \Psi^\dagger (x) (-i\sigma_z\partial_x)\Psi(x)+\right.\nonumber\\
\left.\chi^\dagger (x)\,\,\,\,(i\sigma_z\partial_x) \chi(x)\right],
\end{eqnarray}
where, as before, $\Psi(x)$ and $\chi(x)$ are the Fermionic operators on the two edges. Moreover, the magnetic barrier is now described by the Hamiltonian
\begin{equation}
H_{M_3}=M\int_{-L}^0 dx \left[\Psi^\dagger(x)\sigma_x \Psi^\dagger(x)+\chi^\dagger(x)\sigma_x\chi(x)\right],
\end{equation}
where $M$ again parametrizes the strength of the magnetic barrier.\\
We first analyse the regime where the fractional charge is properly created, that is $t_0(L_2-L_1)/v_F, ML/v_F\rightarrow\infty$. It is easy to understand by intuitive arguments that, in this regime, a quantum dot is formed between the ferromagnetic barrier and the quantum point contact: Imagine to start with a right mover (with, say, spin up), located between the magnetic barrier and the quantum point contact, in the upper edge of the topological insulator. When arriving at the quantum point contact, the electron is converted into a left mover belonging to the other edge (again with spin up). After the backscattering, the particle is a left mover with spin up in the lower edge. When it scatters onto the magnetic barrier, the electron gets reflected in the same edge and flips its spin. When getting back to the quantum point contact, it is backscattered into the upper branch and preserves its spin. The cycle is hence closed when the electron is reflected by the magnetic barrier for the second time. This series of processes is depicted in Fig.3$(a)$. The presence of a semi-integer charge can be understood by means of an intuitive argument as well. A scheme is shown in Fig.3, panels $(b)$ and $(c)$. The gap induced by the quantum point contact allows us to imagine the left/right movers in the upper branch as being continuously attached to the right/left movers in the lower branch. The effective system obtained can then be unfolded and is equivalent to a system with two magnetic barriers with opposite magnetization direction. A fractional charge 1/2 of Jackiw-Rebbi type is hence trapped in the system.\\
\begin{figure}
	\begin{center}
		\includegraphics[width=0.8\linewidth]{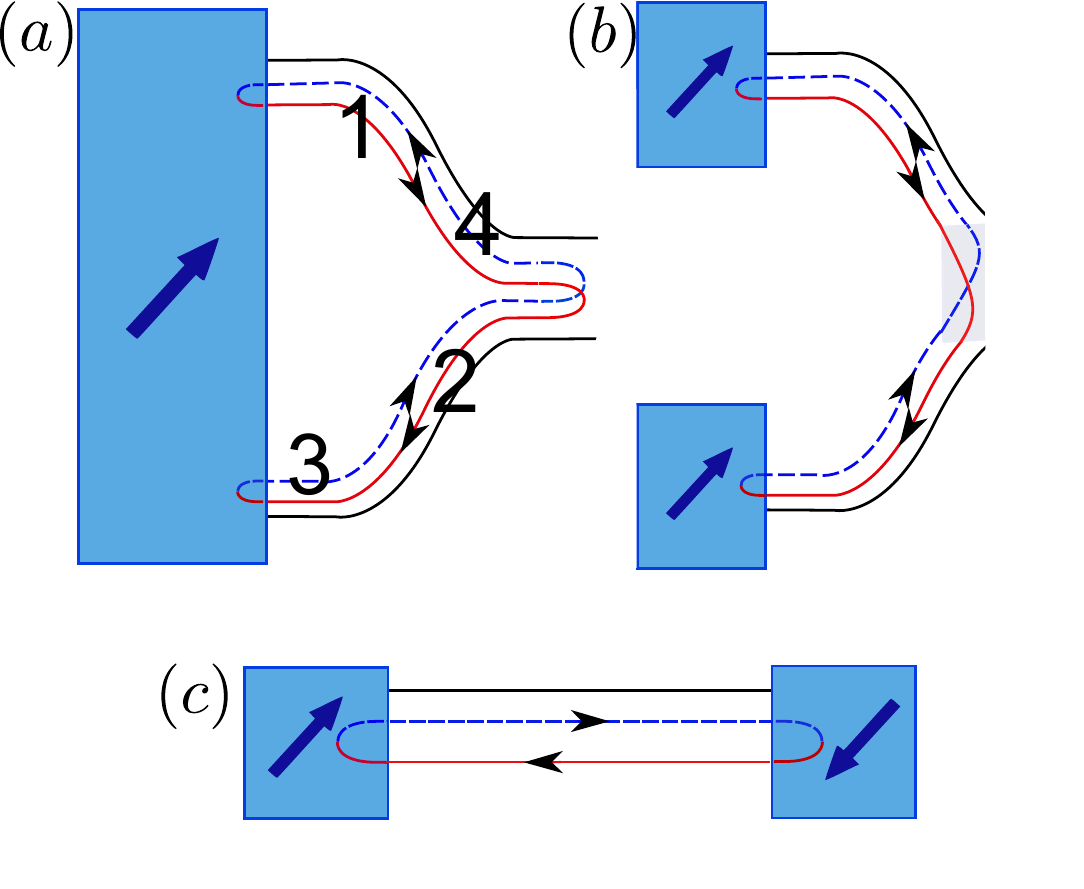}
		\caption{$(a)$ Representation of the scattering mechanism responsible for the existence of the quantum dot. $(b)$ and $(c)$ Unfolding of the system pointing in the direction of the presence of the fractional charge.}
		\label{fig:e_B_mu}
	\end{center}
\end{figure}
To put the discussion on a more formal basis, we define the wavefunction $\psi(x)$ and its four components $\psi_n^{(j)}(x)$, with $j=1,..,4$. We fix the basis in such a way that the first two components of $\psi_n(x)$ represent spin up and spin down electrons in the upper edge, respectively, and the third and fourth components represent spin up and spin down electrons in the lower edge. Then, the Schr{\"o}dinger equation related to $H_{3}$, in the limit of infinite $M_3$ and $t_0$, becomes equivalent, as far as the solutions between $x=0$ and $x=L_1$ are concerned, to
\begin{equation}
H_\infty \psi_n(x)=\epsilon_n\psi_n(x),
\end{equation}
where
\begin{eqnarray}
H_\infty=\lim_{LM,t_0\rightarrow\infty}\left(-iv_F\tau_z\otimes \sigma_z+\right.\nonumber\\
\left. ML \delta(x)I_{2\mathrm{x}2}\otimes \sigma_x+t\delta(x-d)\tau_x\otimes I_{2\mathrm{x}2}\right).
\end{eqnarray} 
Here, $\tau_x$, $\tau_y$, and $\tau_z$ are Pauli matrices in the usual representation, acting on the subspace defined by the existence of both the upper and the lower edge.\\
The Schr{\"o}dinger equation can be solved exactly with the method described in Ref.\cite{timm}. The eigenenergies $\epsilon_n$ are given by $\epsilon_n=v_F k_n$, and the quantized allowed values for $k_n$ are given by
\begin{equation}
k_n=\frac{n\pi}{2L_1},\,\,\,\,\,\,\,n\in {Z}.
\end{equation}
The main points to be noticed are: The presence of a non-degenerate zero energy state, the fact that the spectrum is symmetric around zero energy, and the unusual quantization of a wavevector. The combination of the first two points implies the presence of half integer charge, while the third implies that the effective length of the quantum dot is $4L_1$, in accordance with the fact that an electron has to be backscattered four times to get back to its original point (Fig.3$(a)$). The wavefunction $\psi_n(x)$, inside the dot, is given by
\begin{equation}
\psi_{n}(x)\!=\!\frac{1}{\sqrt{4L_1}}\left(\!e^{ik_{n}x}\!,ie^{-ik_{n}x},i(-1)^ne^{-ik_{n}x},(-1)^ne^{ik_{n}x}\!\right)^T.
\end{equation}
The Fermi operator $C(x)$ for electrons in the dot can hence be constructed on the basis of the momentum resolved Fermi operators $c_n$ associated to the wavefunctions $\psi_n(x)$ as
\begin{equation}
C(x)=\sum_{n=-\infty}^\infty \psi_n(x) c_n.
\end{equation}
The charge operator $Q$ in the quantum dot is then evaluated in the usual regularized form relevant for the Dirac equation by means of the commutator
\begin{equation}
Q=\frac{1}{2}\int_0^{L_1} \left[C^\dagger(x),C(x)\right].
\end{equation}
Due to the non-degenerate zero energy mode, $Q$ only has semi-integer eigenvalues, and a fractional charge is hence trapped in the quantum dot\cite{rajaraman}.\\
Adding electron-electron interactions within a Luttinger liquid theory would be, at this stage, straightforward in view of the techniques described in Refs.\cite{timm,giacomo1}. In the weakly interacting regime, the strong spin density oscillations detectable by means of spin sensitive local probe techniques could be measured\cite{giacomo1}. In the strongly interacting regime, the setup proposed would allow for the detection of the fractional Wigner crystal\cite{prl} by means of experiments involving simple local capacitive coupling to external probes, like the ones used in Ref.\cite{sch1,sch2,sch3}.\\
In this Letter, we however focus on the non-interacting picture, relevant for weakly interacting topological insulators, with large dielectric constant \textit{e.g.} Hg(Cd)Te.
\begin{figure}
	\begin{center}
		\includegraphics[width=0.8\linewidth]{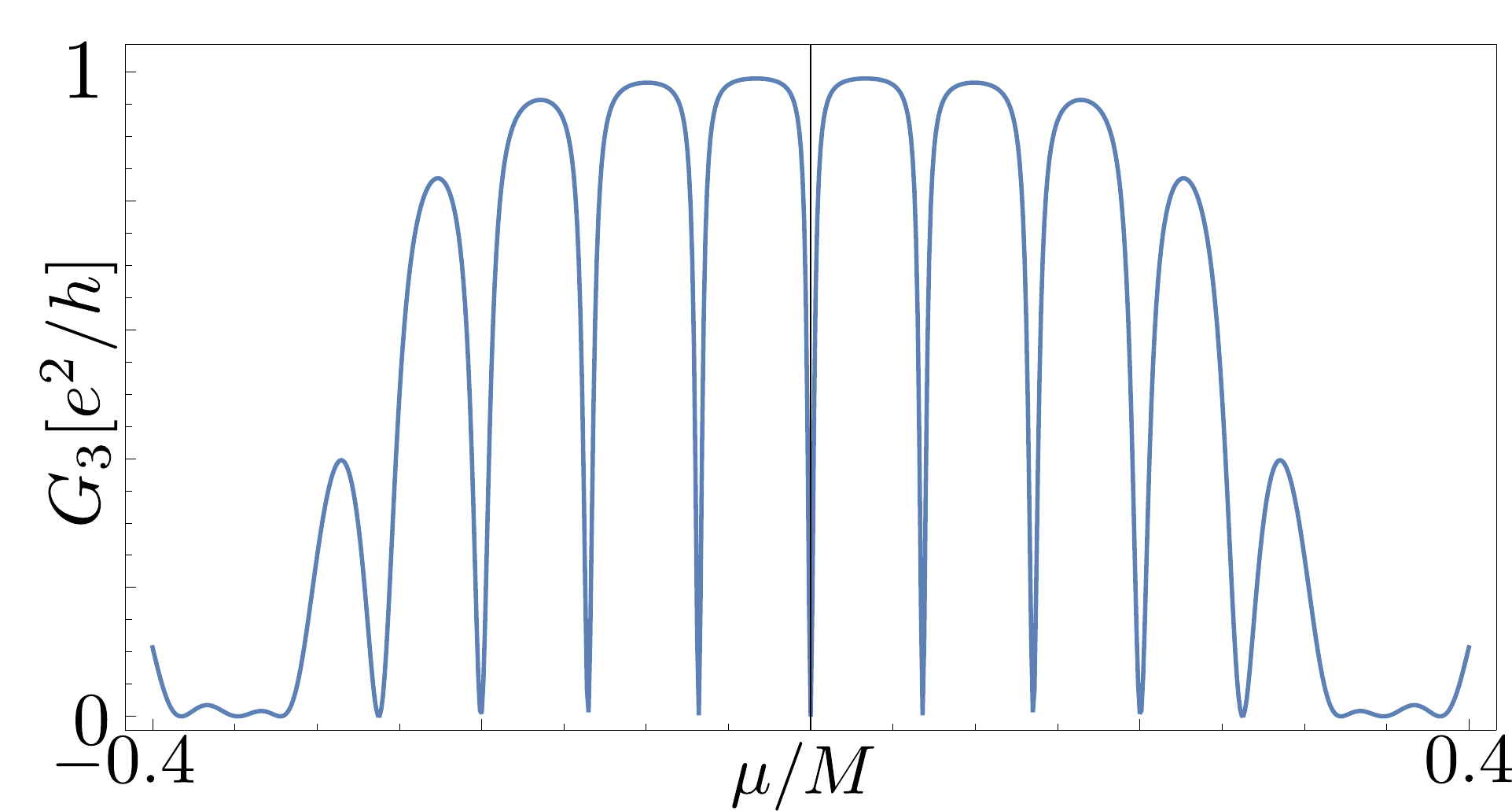}
		\caption{$G_3$ as a function of the gate voltage for $M=2\pi/L$, $L_1=1.25L$ and $L_2=L/2$}
		\label{fig:e_B_mu1}
	\end{center}
\end{figure}
In this regime, we now investigate a finite strength $t_0$ of the backscattering at the point contact and consider the linear conductance $G_3$ when a small bias is applied between the upper and the lower edge. The idea is the following one: In the absence of the point contact, but in the presence of the magnetic barrier, there is no current flowing between the two edges and all incoming particles are reflected. In the presence of the quantum point contact, but in the absence of the magnetic barrier, there is no backscattered current. In the presence of both the magnetic barrier and the quantum point contact, a quantum dot necessarily containing a fractional charge is formed between them. Probing the existence of bound states in the dot is hence equivalent to detecting the fractional charge. Such bound states manifest themselves as dips in $G_3$, since only when the gate voltage puts the energy levels in the dot on resonance for transport, intra-edge backscattering becomes significant. The gate voltage is applied to both the dot and the quantum point contact. The result is shown in Fig.4.  The behaviour of $G_3$ is due to the non-trivial spin structure in the quantum dot, and hence, ultimately, to the presence of the ferromagnetic barrier. Breaking axial spin symmetry and hence introducing a term like $H_s$ in the Hamiltonian of the quantum point contact does not invalidate the reasoning given in this section, since it only introduces a new forward scattering mechanism.
\section{Conclusion}
In this Letter, we have addressed the realization and the detection of fractional solitons in two-dimensional topological insulators. We have first pointed out that a setup involving two magnetic barriers and a quantum point contact is capable of detecting fractional solitons. With respect to previous proposals, this setup has the advantage of being suitable for the implementation of isolated fractional solitons, that are interesting for topological pumping experiments, once the tunneling between the edges is switched off. Afterwards, we have presented a modified setup allowing for the implementation and detection of half-integer charges. The latter setup consists of a single magnetic barrier and a point contact. The detection of the fractional charge, in this setup, does not require any time dependence of the magnetization.
\acknowledgments
Financial support by the DFG (SPP1666 and
SFB1170 “ToCoTronics”), the ENB Graduate School
on “Topological Insulators” and the Stuedienstiftung des Deutschen Volkes is acknowledged

\end{document}